\documentclass[twocolumn,showpacs,prl,aps,floatfix,reprint]{revtex4-1}
\usepackage{amssymb}
\usepackage{graphicx}
\usepackage{dcolumn}
\usepackage{amsmath}
\usepackage{amsfonts}
\usepackage{color}
\begin{document}

\title{The neutron-skin thickness of $^{48}$Ca from a nonlocal-dispersive-optical-model analysis}
\author{M. H. Mahzoon$^{1,2,4}$, M. C. Atkinson$^1$,  R. J. Charity$^3$, and W. H. Dickhoff$^1$} 
\affiliation{${}^1$Department of Physics,
Washington University, St. Louis, Missouri 63130, USA}
\affiliation{${}^2$Department of Physics and Astronomy, Michigan State University, East Lansing, MI 48824, USA}
\affiliation{${}^3$Department of Chemistry,
Washington University, St. Louis, Missouri 63130, USA}
\affiliation{${}^4$Department of Physics, School of  Science and Methematics, Truman State University,Kirksville, MO 63501, USA}
\date{\today}

\begin{abstract}
A nonlocal dispersive-optical-model analysis has been carried out for neutrons and protons in $^{48}$Ca. Elastic-scattering angular distributions, total and reaction cross sections, single-particle energies, the neutron and proton numbers, and the charge distribution have been fitted to extract the neutron and proton self-energies both above and below the Fermi energy. From the single-particle propagator resulting from these self-energies, we have determined the charge and neutron matter distributions in $^{48}$Ca.  A best fit neutron skin of 0.249$\pm$0.023~fm is deduced, but values up to 0.33 fm are still consistent.  The energy dependence of the total neutron cross sections is shown to have strong sensitivity to the skin thickness. 
\end{abstract}

\maketitle

A fundamental question in nuclear physics is how the constituent neutrons and protons are distributed in the nucleus. In particular, 
for a nucleus which has a large excess of neutrons over protons, are the extra neutrons distributed evenly over the nuclear volume 
or is this excess localized in  the periphery of the nucleus forming a neutron skin? 
A quantitative measure is provided by the neutron-skin thickness $\Delta r_{np}$ defined as the difference between neutron and proton rms radii, \textit{i.e.}, $\Delta r_{np} = r_{n} - r_{p}$.

The nuclear symmetry energy which characterizes the variation of the binding energy as a function of neutron-proton asymmetry, opposes the creation of nuclear matter with excesses of either type of nucleon.
The extent of the neutron skin is determined by the relative strengths of the symmetry energy between the central near-saturation 
and peripheral less-dense regions.  Therefore $\Delta r_{np}$ is a measure of the density dependence of the symmetry energy around saturation~\cite{Typel01,Furnstahl02,Steiner05,RocaMaza11}. This dependence is  very important  for determining many nuclear properties, including masses, radii, and the location of the drip lines in the chart of nuclides. Its importance extends to astrophysics for understanding supernovae and neutron stars~\cite{Horowitz01,Steiner10}, and to heavy-ion reactions~\cite{li08}.  

Given the importance of the neutron-skin thickness in these various areas of research, a large number of studies (both  experimental and theoretical) have been devoted to it~\cite{Tsang12}. While the value of $r_p$ can be determined quite accurately from electron scattering~\cite{Angeli:2013}, the 
 experimental determinations of $r_n$  are typically model dependent~\cite{Tsang12}. However, the use of parity-violating electron scattering does allow for a nearly model-independent extraction of this quantity~\cite{Horowitz98}. The present value for $^{208}$Pb extracted with this method from the PREX collaboration yields a skin thickness of  $\Delta r_{np}$=0.33$^{+0.16}_{-0.18}$~fm \cite{PREX12}. Future electron-scattering measurements are expected to reduce the experimental uncertainty.

In this work we present an alternative method of determining $r_n$ using a dispersive-optical-model (DOM) analysis of bound and scattering data to constrain the nucleon self-energy  $\Sigma_{\ell j}$.
This self-energy is a complex and nonlocal potential that unites the nuclear structure and reaction domains~\cite{Mahaux91,Mahzoon:2014}.
 The DOM was originally developed by Mahaux and Sartor~\cite{Mahaux91}, employing local real and imaginary potentials connected through dispersion relations. However, only
with the introduction of nonlocality can realistic self-energies be obtained~\cite{Mahzoon:2014,Dickhoff:2017}. 
The Dyson equation then determines the single-particle propagator or Green's function $G_{\ell j}(r,r';E)$ from which bound-state and scattering observables can be deduced. In particular the particle number and density distributions of the nucleons can be inferred, thus enabling us to probe the neutron skin of a nucleus.

We recently extracted the proton and neutron self-energies in the symmetric $^{40}$Ca system~\cite{Mahzoon:2014}. A functional form of the self-energy was assumed which was based on 
theoretical expectations~\cite{Waldecker:2011,Dussan:2011} and the long history of fitting elastic-scattering data. This study allowed for the spectral strengths of proton and neutron orbitals to be calculated both below and above the Fermi energy~\cite{Mahzoon:2014,Dussan:2014}. We have now extended this work to include $^{48}$Ca allowing the asymmetry dependence of these spectral strengths to be determined and the neutron skin to be extracted. Both proton and neutron self-energies have been determined, but this work will concentrate mostly on the neutron self-energy as $r_p$ is known to high precision from electron scattering. Some relevant proton results will also be presented.    

The point neutron or proton density distributions are given as a sum over contributions from each $\ell j$ orbit,\textit{ i.e}
\begin{equation}
\rho(r) = \frac{1}{4\pi} \sum_{\ell,j} \left( 2 j +1 \right) n_{\ell j}(r,r)
\end{equation}
obtained from the one-body density matrix
\begin{equation}
n_{\ell j}(r,r') = \int_{-\infty}^{\varepsilon_{F}} dE \, S_{\ell j}(r,r';E) ,
\end{equation}
with the spectral density given by
\begin{equation}
   S_{\ell j}(r,r';E) =  \frac{1}{\pi}\textrm{Im} \left [G_{\ell j}(r,r';E) \right].
\end{equation}

Before considering the neutron skin for $^{48}$Ca, it is important to benchmark this method by considering the predictions for the $N=Z$ system $^{40}$Ca. In our previous work \cite{Mahzoon:2014},  the neutron and proton self-energies were assumed identical apart from the Coulomb  
contribution and they were fit simultaneously to a large amount of data including the charge distribution. 
It is therefore not surprising that the point neutron density distribution was very similar to the proton one. The extracted skin thickness is $\Delta r_{np}$=-0.06~fm where we have used the experimental value for the proton rms radius~\cite{Angeli:2013} as a reference. Indeed, a very small, but negative, value is expected as protons have an extra repulsion from the Coulomb force which forces them slightly further apart. Theoretical predictions for this system range from $\Delta r_{np}$=-0.02 to -0.10~fm~\cite{Agrawal05,Avancini07,Centelles09}, consistent with our result. 

Given that the fitted neutron and proton self-energies are identical apart from the Coulomb potential, our extracted result may be considered highly constrained. We have therefore refit just the neutron data alone to see if this value changes. At the same time, it is also important to obtain an error estimate arising from the uncertainties of the experimental data.
The statistical uncertainties associated with the fitted scattering data sets are typically quite small, but the largest uncertainties are systematic associated with the normalization of the cross sections. 
In addition, the large number of elastic differential cross sections in our data sets overwhelmed the total calculated $\chi^2$, giving little sensitivity in the fits to total and reaction cross sections and bound-state data. We therefore implemented a weighted $\chi^2$ fit giving more weight to these other data sets so they properly influence the final outcome. It is thus clear that we cannot use the standard $\chi^2$ analysis which assumes all errors are statistical to estimate the $\Delta r_{np}$ error estimates. Instead we followed Varner et al.~\cite{Varner91}, who in their global optical-model fits used a bootstrap method. New modified data sets were created from the original data by randomly renormalizing each angular distribution or excitation function within $\pm$10\% to incorporate fluctuations from the systematic errors. Forty such modified neutron data sets were generated and refit. The mean of the new fitted skin thickness is $\Delta r_{np}$=-0.065$\pm$0.008~fm. This is almost identical to the original value obtained from fitting the combined neutron and proton data.  

We now return to $^{48}$Ca. The neutron and proton elastic-scattering angular distributions, total and reaction cross sections, and the single-particle level data used in this case are the same as from our previous local DOM fits~\cite{Mueller:2011}.  In addition  the experimental charge distribution~\cite{deVries:1987} is included in the fit and calculations are constrained to give correct total numbers of neutrons and protons. 
The parametrization of the self-energy used in these fits is similar to that used previously for $^{40}$Ca~\cite{Mahzoon:2014} except neutron-proton asymmetry terms [($N-Z)/A$] have been added to the various real and imaginary components. Some parameters are left fixed at the values used for $^{40}$Ca. Details of the parametrization can be found in Ref.~\cite{sup}.

Briefly, the real part of the self-energy is comprised of local Coulomb and spin-orbit contributions plus a nonlocal Hartree-Fock potential. 
The imaginary potential has two nonlocal components, one surface localized to capture the contributions from 
long-range correlations and the other spread out over the volume of the nucleus to account for short-range correlations.  Except for the Coulomb potential, which is derived from the experimental charge distribution, the  radial dependence of these contributions are parametrized with Woods-Saxon or derivatives of Woods-Saxon form factors.  The magnitudes and radius parameters of these Woods-Saxon terms for the asymmetry contributions to the Hartree-Fock, volume, and surface imaginary contributions are varied independently in the $^{48}$Ca fits. The asymmetric Hartree-Fock component is allowed to have a different nonlocality parameter and for the main $N=Z$ component, the radius parameter is also allowed to vary. Gaussian nonlocality~\cite{Perey:1962} is assumed for nonlocal terms. 
 The energy dependence of the asymmetric volume-imaginary potential is kept fixed to the $^{40}$Ca result, but  for the asymmetric surface-imaginary contribution, many aspects are allowed to vary for both protons and neutrons, including its magnitude, energy dependence, and nonlocality parameters. Dispersion relations are enforced between the dynamic real and imaginary components. 

\begin{figure}[tbp]
\begin{minipage}{\columnwidth}
   \makebox[\columnwidth]{
      \includegraphics{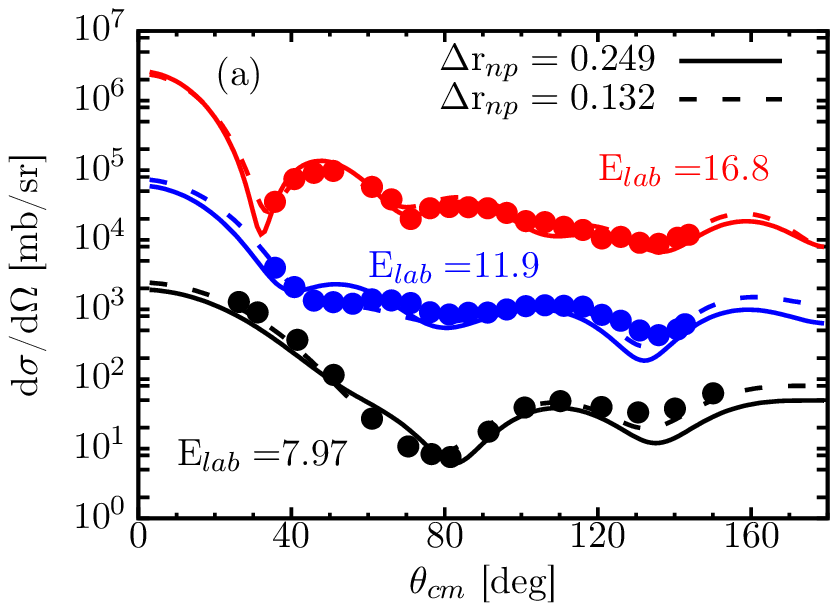}
   }
   \vspace{-0.7cm}
\end{minipage}
\begin{minipage}{\columnwidth}
   \makebox[\columnwidth]{
      \includegraphics{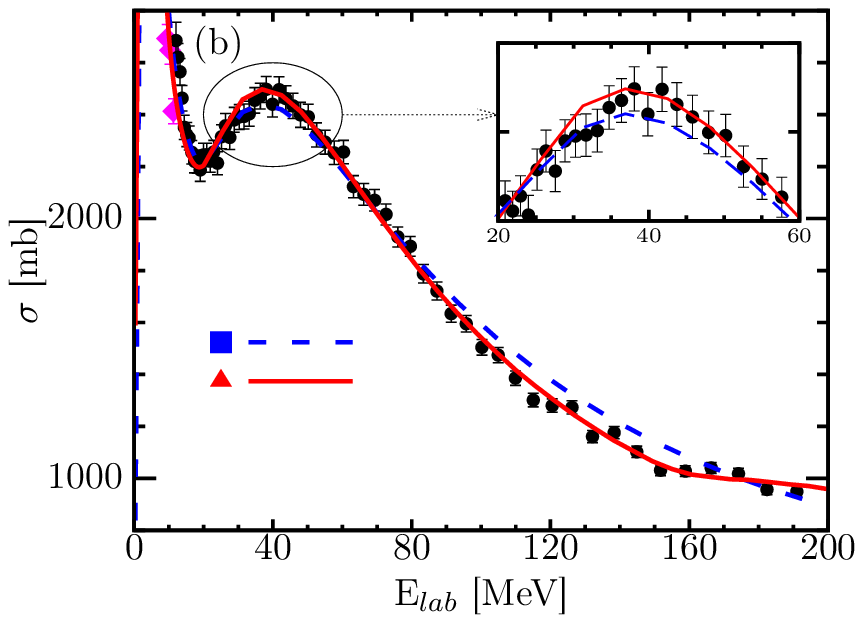}
   }
   \vspace{-0.7cm}
\end{minipage}
\begin{minipage}{\columnwidth}
   \makebox[\columnwidth]{
      \includegraphics{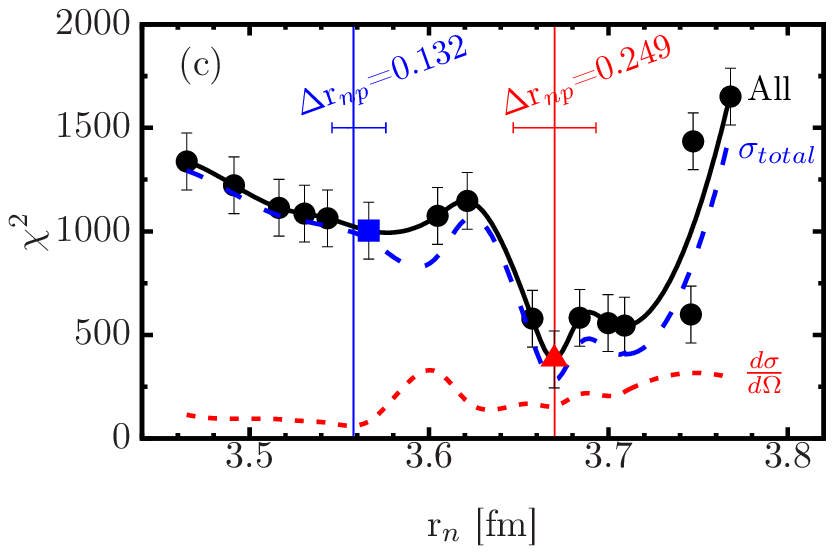}
   }
   \vspace{-0.6cm}
\end{minipage}
\caption{
   (Color online) (a) Comparison of experimental $n$+$^{48}$Ca elastic-scattering angular distributions \cite{Hicks:1990,Mueller:2011} to the best DOM fit of all data (solid curves) and to a constrained fit with the skin thickness forced to $\Delta r_{np}$=0.132~fm (dashed curves) which is consistent with the \textit{ab initio} result. The higher-energy data and calculations have been offset along the vertical axis for clarity.   (b) Comparison of the experimental total neutron cross sections of $^{48}$Ca (diamonds \cite{Harvey:1985}, circles \cite{Shane:2010}) to DOM fits with constrained values of $r_{n}$. The curve labeled with a triangle is for the $r_{n}$ value of our best fit, while the curve labeled with a square is for a value consistent with the \textit{ab initio} result (see Fig. 1(c)). (c) The $\chi^2$ from fitting all data (solid curve) and its contribution from fitting the elastic-scattering angular distributions and total neutron cross section (short-dashed and long-dashed curves respectively). Each data point corresponds to an average of fitted values with very similar $r_{n}$ values.} 
\label{fig:three}
\end{figure}

For neutrons, the fitted elastic-scattering angular distributions and total cross sections are displayed as the solid curves in Figs.~\ref{fig:three}(a) and \ref{fig:three}(b), respectively. The fit parameters can be found in Ref.~\cite{sup}. The fits to the neutron elastic-scattering angular distributions in Fig.~\ref{fig:three}(a) are now better than those we obtained from our previous local DOM fits in Ref.~\cite{Mueller:2011}.   
%The fitted Hartree-Fock component is shallower and narrower for neutrons compared to protons. 
Both the protons and neutrons show enhanced surface absorptive potentials relative to those found for $^{40}$Ca, with a particularly strong enhancement for protons below the Fermi energy. 
More details of the fits to data can be found in Ref.~\cite{sup}.

\begin{figure}[tbp]
   \begin{minipage}{\columnwidth}
      \makebox[\columnwidth]{
         \includegraphics{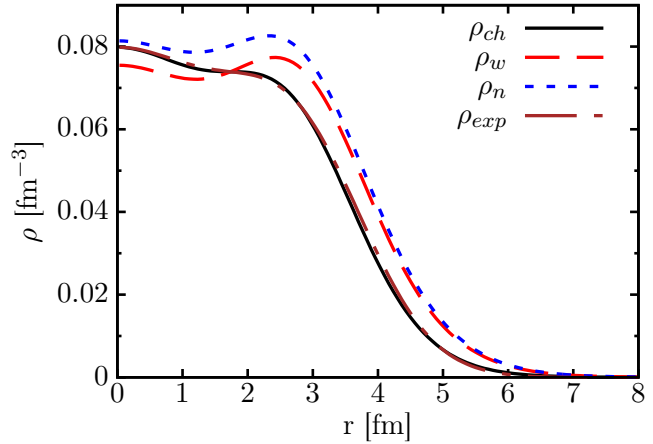}
      }
   \end{minipage}
   \caption{(Color online) Comparison of experimental ($\rho_{exp}$) and fitted ($\rho_{ch}$) charge distribution for $^{48}$Ca. The neutron matter distribution is plotted as $\rho_{n}$, while the weak charge distribution is plotted as $\rho_{w}$.}
   \label{fig:dist}
\end{figure}

The experimental charge distribution was well reproduced as can be seen in Fig.~\ref{fig:dist}. The neutron matter distribution clearly extends out to larger radii forming a neutron skin. The weak charge distribution, calculated from the fitted neutron and point proton distributions~\cite{PREX12}, is shown as $\rho_w$.  
The neutron-skin thickness of $^{48}$Ca deduced from these distributions is $\Delta r_{np}$=0.249$\pm$0.023~fm where we again used the bootstrap method to estimate the experimental uncertainty. 
This value overlaps with the range of values (0.12-0.26 fm) predicted with 48 reasonable nuclear energy-density functionals in Ref.~\cite{Piekarewicz:2012} but is large compared to the range of 0.12-0.15~fm obtained with the  \textit{ab initio} coupled-cluster method~\cite{Hagen:2016}.

To further understand which data in the fits exhibit the most sensitivity to skin thickness, we have made constrained fits where selected values of $r_n$ are forced in the DOM calculations. 
This is achieved by varying the radius parameters of the main real potential ($r^{HF}_{n}$ and $r^{HFasy}_{n}$ in Ref.~\cite{sup}) and refitting the other asymmetry dependent parameters.  Our weighted $\chi^2$ as a function 
of the calculated $r_n$ is plotted as the data points in Fig.~\ref{fig:three}(c) and the 
absolute minimum at $r_n$=3.67~fm corresponds to our skin thickness of 0.249~fm.
We found some fine-scale jitter in the variation of $\chi^2$ with $r_n$, and 
because we want to concentrate on the larger-scale variation, the data points shown in Fig.~\ref{fig:three}(c) are local averages with the error bars giving the range of the jitter.    

The location of the \textit{ab initio} results is also indicated at $r_{n} \sim $3.56~fm where the $\chi^2$ is larger. We have subdivided this $\chi^2$ into its contributions from its two most important components (dashed curves); from  the elastic-scattering angular distributions and from the total neutron cross sections. The former has a smaller sensitivity to $r_n$ and its $\chi^2$ is slightly lower for the smaller values of $r_n$ which are more consistent with the \textit{ab initio} result as illustrated in Fig.~\ref{fig:three}(a) where a fit with a forced value of $\Delta r_{np}$=0.132, is compared to our best fit and to the data. While this new calculation improves the 
reproduction of these data, the deviations of both curves from the data are typical of what one sees in global optical-model fits. In addition, these experimental angular distributions only cover a small range of bombarding energies (7.97 to 16.8 MeV) and may not be typical of other energies.

The total cross section exhibits larger sensitivity and the experimental data cover a large range of neutron energies (6 to 200~MeV). Two data sets are available (circles and diamonds) but are inconsistent by $\sim$10\% at $E_{lab}\sim$10 MeV where their ranges overlap. We consider the high-energy data set~\cite{Shane:2010} (circles) more accurate as it was obtained with $^{48}$Ca metal, while the low-energy set~\cite{Harvey:1985} (diamonds) employed $^{48}$CaCO$_{3}$ and required a subtraction of $\sim$70\% of the signal due to neutron absorption from the CO$_{3}$ component. Therefore we have chosen to display the $\chi^2$ contribution only from the high-energy set.  This $\chi^2$ exhibits a broad minimum from $r_n$= 3.66 to 3.75~fm allowing values of $\Delta r_{np}$ up to 0.33~fm.  

Figure~\ref{fig:three}(b) illustrates the sensitivity to $r_n$ where the solid and dashed curves correspond to the fits indicated by triangular and square data points in Fig.\ref{fig:three}(c), respectively. The former is the best fit while the latter has a skin-thickness consistent the \textit{ab initio} result. The latter calculation under predicts the maximum at 40 MeV while over predicting the 80-180~MeV region. These differences arise almost exclusively from the elastic-scattering contribution to the total cross section whose energy dependence displays large-scale oscillations due to the interference between transmitted and externally scattered neutrons~\cite{Satchler:1983} leading to a phase shift that depends on the size and depth of the real component of the neutron self-energy.    

\begin{figure}[tbp]
   \begin{minipage}{\columnwidth}
      \makebox[\columnwidth]{
         \includegraphics{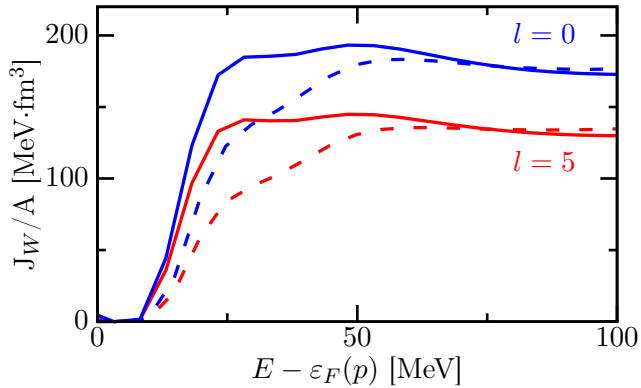}
      }
   \end{minipage}
   \caption{(Color online) Comparison of the integrated imaginary potential for 
      protons on $^{40}$Ca (dashed curves) and $^{48}$Ca (solid curves) obtained from our DOM fits. Results are given for the two indicated $\ell$ values. }
      \label{fig:protons}
   \end{figure}
   As the proton-neutron interaction is stronger than its neutron-neutron counterpart, the imaginary, or absorptive, part of the proton self-energy should be quite sensitive to the neutron density distribution.  In standard local optical-model fitting of scattering data, it has been long known that 
   integrated potentials are well constrained even though there can be some ambiguities in the fit parameters~\cite{Mahaux91}. For our nonlocal self-energy, such volume integrals can be calculated according to 
   \begin{equation}
      J^{\ell}_{W}(E) = 4\pi \int_{0}^{\infty} \!\! dr r^2 \int_{0}^{\infty} \!\! dr' r'^2\ \textrm{Im}\left [ \Sigma_{\ell}(r,r';E) \right ]
   \end{equation} 
   where the nonlocal self-energy is projected onto good total angular momentum and averaged over spin-orbit partners~\cite{Waldecker:2011}.
   Figure~\ref{fig:protons} compares  $J^{\ell}_{W}(E)$ determined from our $p$+$^{48}$Ca fits to those obtained from $p$+$^{40}$Ca Ref.~\cite{Mahzoon:2014}. Results are shown for two representative values of the orbital angular momentum ($\ell$ = 0, 5). The decrease of $J^{\ell}_{W}$ with $\ell$ is a consequence of nonlocality of the imaginary potential which is essential to obtain correct particle number~\cite{Mahzoon:2014} and sum rules~\cite{Dussan:2014}.  
   For energies below $E\sim$50~MeV the absorption from the elastic channel is dominated by surface interactions, and here we see a big increase for $^{48}$Ca as would be expected from a neutron skin. 
   On the other hand, at larger energies we are most sensitive to the interior of the nucleus where the results for $^{40}$Ca and $^{48}$Ca are practically identical. This suggests that the interiors of $^{40}$Ca and $^{48}$Ca are similar and thus the extra neutrons for $^{48}$Ca will develop a skin.

   \begin{figure}[tbp]
      \begin{minipage}{\columnwidth}
         \makebox[\columnwidth]{
            \includegraphics{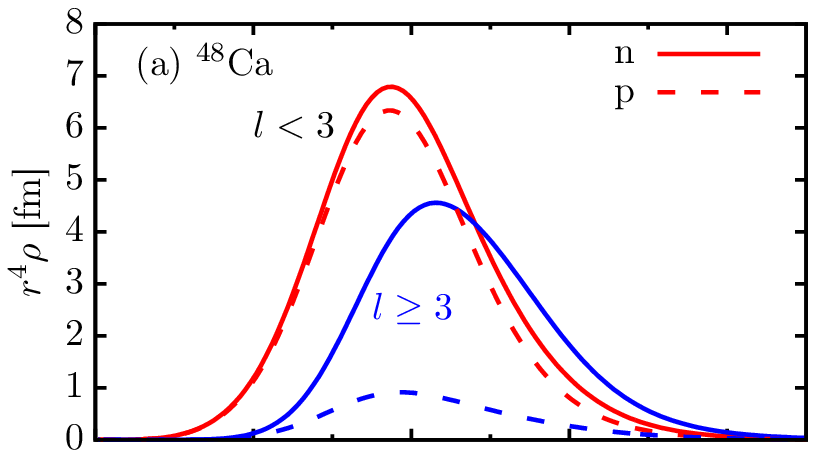}
         }
         \vspace{-0.9cm}
      \end{minipage}
      \begin{minipage}{\columnwidth}
         \makebox[\columnwidth]{
            \includegraphics{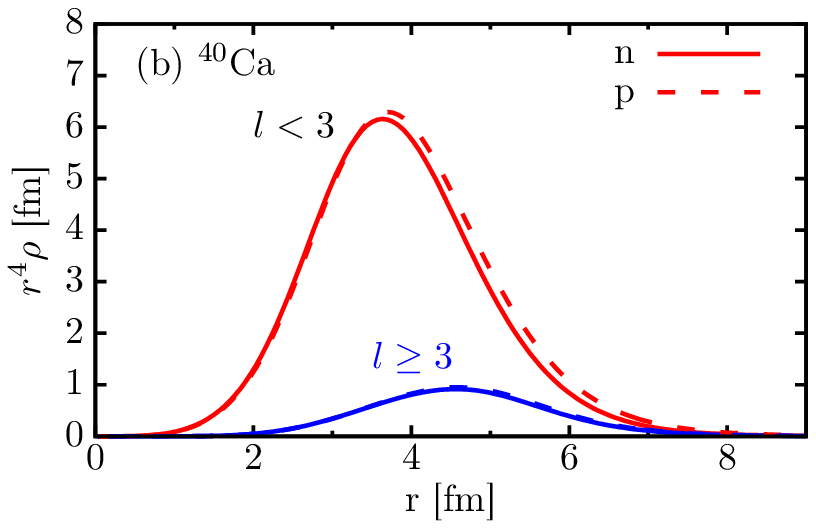}
         }
         \vspace{-0.5cm}
      \end{minipage}
      \caption{(Color online) Decomposition of the $r^4$ weighted point densities for protons and neutrons in (a) $^{48}$Ca and (b) $^{40}$Ca. These are subdivided into the contribution from the lower-$\ell$ orbitals [$s_{1/2}$,$p_{3/2}$, $p_{1/2}$, $d_{5/2}$, and $d_{3/2}$] designated by ``$\ell<3$'' and the remaining higher-$\ell$ orbitals ``$\ell\geq 3$''.}
      \label{fig:density}
   \end{figure}

   To further visualize where the extra neutrons in $^{48}$Ca are located, we show in Fig.~\ref{fig:density} the calculated proton and neutron point distributions weighted by $r^4$ for both $^{40}$Ca and $^{48}$Ca. The rms radii are determined from the integrals of these quantities. These distributions have been subdivided into the contribution from lower-$\ell$ orbitals ($s_{1/2}$, $p_{3/2}$, $p_{1/2}$, $d_{5/2}$, and $d_{3/2}$)  and that from the remaining higher-$\ell$ orbitals which is dominated by the $f_{7/2}$ component. For $^{40}$Ca, the proton and neutron distributions are very similar as expected given there is essentially no neutron skin. For $^{48}$Ca, the contribution from the lower-$\ell$ orbitals, common to both neutrons and protons, is very similar to the $^{40}$Ca results. Not surprisingly, the magnitude of the neutron skin comes predominately from the $f_{7/2}$ orbital, reflecting its centrifugal barrier.

   In conclusion we have performed a nonlocal-dispersive-optical-model analysis of neutron and proton data associated with $^{48}$Ca. We have fitted elastic-scattering angular distributions, absorption and total cross sections,  single-particle energies, and the proton charge distribution while constraining the nucleon numbers. These data are best fit with a  neutron-skin thickness of  
   0.249$\pm$ 0.023~fm, but larger values also give acceptable reproductions of these data. 
 A recent analysis of $(p,n)$ charge-exchange scattering data also points to the possibility of a larger neutron skin~\cite{Pawel17}.
   These skin thicknesses are large compared to recent \textit{ab initio} calculations of 0.12-0.15~fm from~\cite{Hagen:2016}. This disagreement further strengthens the argument for a parity-violating-electron-scattering measurement of this nucleus. We have shown that total neutron cross sections, measured over a large range of neutron energies, exhibit a strong sensitivity to the magnitude of the neutron skin and would permit one to map out the magnitude of this skin for many other stable isotopes in the future.   

   This work was supported by the U.S. Department of Energy, Division of
   Nuclear Physics under grant No. DE-FG02-87ER-40316 and by the U.S. National Science Foundation under grants PHY-1304242, PHY-1613362, and PHY-1520971.

   \end{document}